\documentclass[10pt,twocolumn,showpacs,aps,amsfonts,floatfix,prx]{revtex4-1}
\usepackage{graphicx}
\usepackage{times}
\usepackage{color}
\usepackage[T1]{fontenc}
\usepackage{amssymb}
\newcommand{\be}{\begin{equation}}
\newcommand{\ee}{\end{equation}}
\newcommand{\br}{\begin{eqnarray}}
\newcommand{\er}{\end{eqnarray}}

\newcommand{\bd}{\begin{displaymath}}
\newcommand{\ed}{\end{displaymath}}

\newcommand{\bfig}{\begin{figure}}
\newcommand{\efig}{\end{figure}}

\def\3cdot{\cdot \cdot \cdot}

\def\om0{\omega _0}
\def\Om0{\Omega _0}

\def\text#1{{\rm{#1}}}

\def\->{\rightarrow}
\def\=>{\Rightarrow}
\def\-->{\longrightarrow}
\def\==>{\Longrightarrow}

\def\dag{\dagger}

\def\pr{^\prime}
\def\pr2{^{\prime\prime}}

\def\bfig{\begin{figure}}
\def\efig{\end{figure}}

\begin{document}
\title{Capacitive Coupling of Two Transmission Line Resonators Mediated by the
Phonon Number of a Nanoelectromechanical Oscillator}
\author{O. P. de S\'a Neto$^{1,2}$\email{olimpioqedc@gmail.com}, M. C. de Oliveira$^{2}$\email{marcos@ifi.unicamp.br}
, F. Nicacio$^2$,  G. J. Milburn$^3$}

\affiliation{$^1$ Coordenação de Ciência da Computação,
Universidade Estadual do Piau\'i, CEP: 64202220, Parnaíba, Piau\'i, Brazil.}

\affiliation{$^2$ Instituto de F\'\i sica ``Gleb Wataghin'',
 Universidade Estadual de Campinas, 13083-970, Campinas, S\~ao Paulo, Brazil.}
\affiliation{
$^3$ Centre of Excellence in Engineered Quantum Systems,
University of Queensland, QLD 4072, Brisbane, Australia.}

\date{\today}

\begin{abstract}
Detection of quantum features in mechanical systems at the nanoscale constitutes a challenging task, given the weak interaction with other elements and the available technics. Here we describe how the interaction between two 
monomodal transmission-line resonators (TLRs) mediated by vibrations of a nano-electromechanical oscillator can be described. 
This scheme is then employed for quantum non-demolition detection of the number of phonons in the nano-electromechanical 
oscillator through a direct current measurement in the output of one of the TLRs. For that to be possible an undepleted field inside one of the TLR works as a amplifier for the interaction between the mechanical resonator and the remaining TLR.  We also show how how the non-classical nature of this system can be used for generation of tripartite entanglement and conditioned mechanical coherent superposition states, which may be further explored for detection processes.
\end{abstract}

\maketitle 

\section{Introduction}

The nature of the movement of tiny electromechanical oscillators has proved very intriguing,
receiving attention since the early years of the 
quantum theory \cite{Bohr1958}.
Recently, several groups have been able to 
engineer nano-electromechanical systems (NEMSs) 
with oscillation frequencies up to the GHz scale, despite the challenge to 
sensitively detect movement at that small scale \cite{Knobel2003,LaHaye2004,Naik2006,LaHaye2009,Teufel2009}. 
Indeed it was demonstrated recently that one can cool down to almost the ground state of the mechanical oscillator \cite{o2010quantum,Chan2011,seletskiy2012cryogenic} and to implement experiments near the zero point motion \cite{hertzberg2009back}, therefore inside the quantum regime.
Single electron transistors devices are the natural choice for movement detection, 
but recently electrical transducers of motion using  
circuit quantum electrodynamical devices have been considered \cite{Wei2006,Zhou2006}.
 Indeed, it is interesting to explore the possibilities that a transduction and coupling to other circuit elements may offer for detection purposes.
 
In this article we show that a direct capacitive 
coupling between a mechanical oscillator and two transmission-line resonators (TLR)  
\cite{Milburn2007} enables a quadratic coupling between the TLRs and the mechanical displacement. 
This procedure leads to an efficient method for measurement of the mean phonon number of 
the mechanical oscillator by current measurements on the 
device. The coupling between the mechanical oscillator and the radiation field inside the resonators is amplified in an undepleted regime, allowing a direct quantum non-demolition measurement (QND) of the mechanical resonator mean number of phonons in a simple setup. As a secondary result, given the nature of the interaction between the elements, an entangled state can be generated between the TLR modes and the mechanical resonator states, which might be useful for further application in quantum information processing, or for detection purposes.

\section{Model}\label{model}

\begin{figure}[h!]
{\includegraphics[scale=0.8]{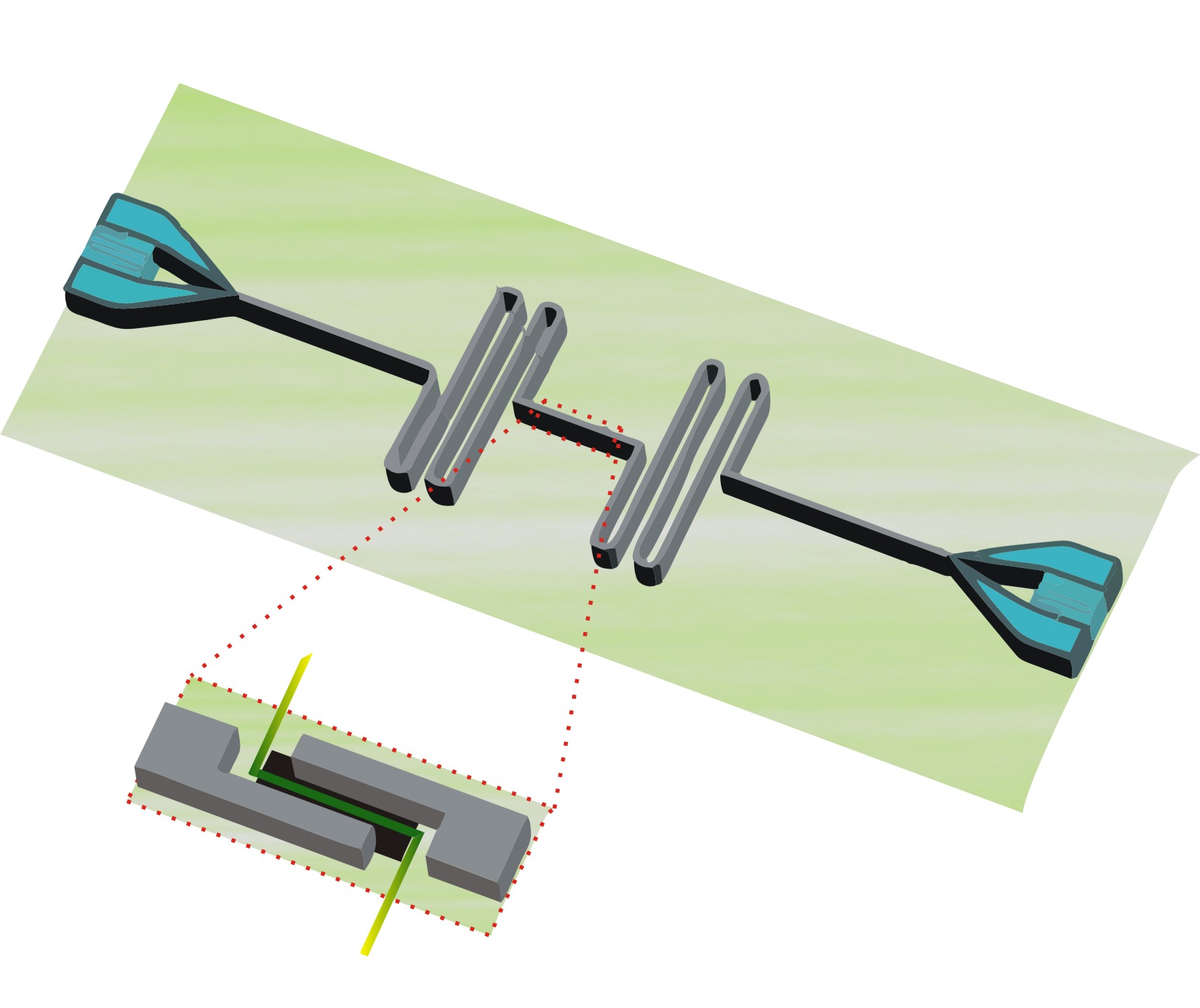}}
\caption{Capacitive coupling of two TLRs mediated 
by an oscillatory NEMS. A double clamped mechanical resonator (green) is eletrically coupled to two transmission line resonators (grey).}
\label{circuit}
\end{figure}

Quantum features for a mechanical oscillator manifest only when its  
oscillation frequency $\nu$ reaches the $\nu>k_BT/\hbar$ limit. 
For typical temperatures of a few $mK$, $\nu$ must be of the order of $GHz$. 
Since $\nu\propto l^{-1}$, where $l$ is a typical dimension of the oscillator, 
this requires $l$ to be of the order of a few $nm$. 
To test the quantum nature of those oscillations constitutes a real challenge. 
A natural way for probing it is through the direct electrical coupling 
of the mechanical oscillator
to radiation at the microwave scale as has been recently 
demonstrated \cite{o2010quantum,hertzberg2009back,teufel2011sideband}. 
In those cases 
standard electrical measurements can be used to monitor 
the mechanical oscillations.
In the same spirit, we consider
two TLRs capacitively coupled to a mechanical oscillator as
depicted in Fig.\ref{circuit}. 
Since the capacitance changes with the distance,
the mechanical oscillations of the NEMS change 
the distributed capacitances of the circuit:  
$C_1(t) = \frac{\epsilon_0A}{\left(d-x(t)\right)}$ between TLR 1 and the NEMS and 
$C_2(t) = \frac{\epsilon_0A}{\left(d+x(t)\right)}$ between the TLR 2 and the NEMS,
where $\epsilon_0$ is the vacuum dielectric constant, $A$ 
is the lateral area of the NEMS and $d$ is the equilibrium distance of both TLRs
from the NEMS, here assumed to be equal. Also $x(t)$ is the time dependent displacement of the NEMS from its center of mass.
At the NEM's equilibrium position, 
we define the equilibrium capacitance as 
$C_{\rm eq} = \epsilon_0A/d$
and we avoid short-circuit stating that $ \max |x(t)| < d$. 

\begin{figure}[h]
{\includegraphics[scale=0.3]{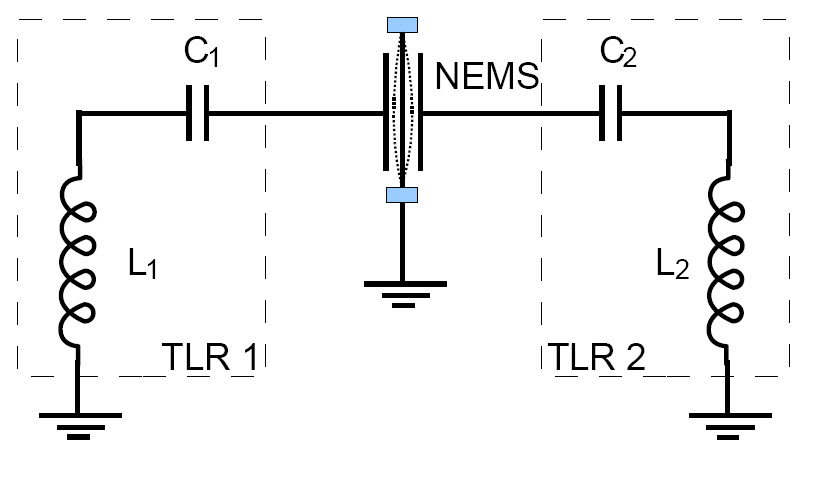}}
\caption{Schematic circuit corresponding to Hamiltonian (\ref{hamil1}), a single mode approximation was considered for both the TLRs and the NEMS.}
\label{circuit2}
\end{figure}

By considering the distributed voltage
and current in the corresponding circuit, after some algebraic manipulations we derive the system Hamiltonian 
(See the appendices for a complete derivation of Hamiltonians (\ref{hamil1}) and (\ref{total})),%
\br {\cal H}&=&\frac{1}{2}\sum_{i=1}^2\left(\frac1{L_i}P_i^2+\frac1{\widetilde{C}_i}Q_i^2\right)
+\frac{\left(d^2-x^2(t)\right)}{2d\epsilon_0 A}Q_1Q_2\nonumber\\
&&-\sum_{i=1}^2\left[\frac{\left((-1)^id+x(t)\right)}{2d}V_{C_{T}}(t)\right]Q_i,\label{hamil1}\er
where  
$Q_i$, $P_i$, $L_i$ are, respectively, 
the charge, the magnetic flux and the impedance at the TLR $i=1,2$. 
We have also defined   $V_{C_T} \equiv V_{C_1}-V_{C_2}={Q_1(t)}{C_1(t)}^{-1}-{Q_2(t)}{C_2(t)}^{-1}$,
and ${\widetilde{C}_i}^{-1}\equiv {C_i}^{-1}+\frac{\left(d^2-x^2(t)\right)}{2d\epsilon_0 A}$, 
where ${C}_i$ is the capacitance in each TLR. In that Hamiltonian we have assumed that only one mode on each TLR is significant in the coupling with the NEMS fundamental mode, as depicted in Fig. (\ref{circuit2}).

Before full quantization of (\ref{hamil1}), 
we will assume a regime of rapid oscillation of the NEMS,
by writing 
\be
x(t) = \sqrt{\frac{\hbar}{2m\nu}}(be^{-i\nu t}+b^\dagger e^{i\nu t}).
\ee
For rapid oscillations ($\nu t>>1$), 
$\left\langle x(t)\right\rangle\approx 0$, furthermore $\left\langle x^2(t)\right\rangle\approx\frac{\hbar}{m\nu}(\left\langle b^\dagger b\right\rangle+\frac{1}{2})\equiv x_{\rm rms}^2$.
Thus a relation between the frequencies  
$\omega_i^2\equiv(L_iC_i)^{-1}$ and $\widetilde{\omega}_i^2\equiv(L_i\widetilde{C}_i)^{-1}$ is obtained as 
\be \label{wtil3}
{\widetilde{\omega}_i^2} = {\omega_i^2}+\frac{\omega_{i,\rm eq}^2}{2}\left(1-\frac{x_{\rm rms}^2}{d^2}\right), 
\ee 
where $\omega_{i,{\rm \rm eq}}^2\equiv{(C_{\rm eq}L_i)}^{-1}$. 
For the sort of device we are looking for, it is reasonable to assume 
$x^2_{\rm rms}/d^2=10^{-6}$  \cite{fabric,Ptoday}
and we disregard it from Eq. (\ref{wtil3}) in a first approximation.
Typical experimental values settle $\omega_i=6$ GHz \cite{fabric}, and if  the second term in (\ref{wtil3}) 
is of the same order it should also be taken into account. We keep the maximal value the 
second term can take, adopting $\widetilde{\omega}_i^2=\omega_i^2+\omega_{i,\rm eq}^2/2$, 
meaning that $ {\widetilde{C}_i}^{-1}={C_i}^{-1}+{(2C_{\rm eq})}^{-1}$.
Assuming 
\be \label{conjug}
Q_j =  \sqrt{ \textstyle{\frac{ \hbar }{2 L_j \widetilde{ \omega}_j} }}(a^\dagger_j+a_j) \,\, {\rm and} \,\, 
P_j = i\sqrt{ \textstyle{\frac{ \hbar L_j \widetilde{\omega}_j }{ 2 }}}(a^\dagger_j-a_j),
\ee
which follow the standard commutation relation we obtain
\begin{eqnarray} 
{H} & = & H_0 + H_{\rm int} + H_{\rm d}, \label{total}
\end{eqnarray}
where 
\begin{eqnarray} 
H_0 &=& \hbar \widetilde{\omega}_1(a_1^\dagger a_1 + \textstyle{\frac12}) +
\hbar \widetilde{\omega}_2(a_2^\dagger a_2+\textstyle{\frac12}),
\end{eqnarray}
$H_{\rm int}$ is the  
coupling between the two TLRs mediated by NEMS phonon number operator,
\begin{eqnarray} 
H_{\rm int} &=&  \frac{\hbar\left[ d^2 - \frac{\hbar}{m\nu} (b^\dagger b + \textstyle{\frac{1}{2}}) \right] }
                  { 4d^2C_{\rm eq} \sqrt{L_1L_2\widetilde{\omega}_1\widetilde{\omega}_2} } 
(a^\dagger_1+a_1) (a^\dagger_2+a_2),\label{hi}
\end{eqnarray}
and $H_{\rm d}$ is the Hamiltonian due to the voltage 
induced by the NEMS oscillations
\begin{eqnarray} 
H_{\rm d} &=& \sqrt{\frac{\hbar}{8}} \, V_{C_T}(t) \left[ \frac{(a^\dagger_1+a_1)}{\sqrt{L_1\widetilde{\omega}_1}} + 
        \frac{(a^\dagger_2+a_2)}{\sqrt{L_2\widetilde{\omega}_2}}\right] .\label{hd}
\end{eqnarray}

Assuming the two TLR fields in resonance, $\widetilde{\omega}_1=\widetilde{\omega}_2=\widetilde{\omega}$, and as a first approximation that $Q_1(t)\approx Q_2(t)$, the Hamiltonian (\ref{hd}) can be neglected.
Also in a referential rotating with  $\widetilde{\omega}$ we can neglect the rapidly oscillating terms, $a_1a_2e^{-2i\widetilde{\omega}t} + {\rm H.c.} $ from Eq.(\ref{hi}), which now reads
\begin{eqnarray}
H^{I}_{\rm int}&=&\hbar\left(\theta_{0}+\theta b^{\dag}b\right)\left(a_{1}^{\dag}a_{2}+a_{1}a_{2}^{\dag}\right),
\label{Hamilf}
\end{eqnarray}
where $\theta_{0} = \frac{ \widetilde{ \omega } \widetilde{C}_1 }{ 4C_{\rm eq} } \left(1 - \frac{\hbar}{2d^2m\nu}\right)\approx \frac{\widetilde{ \omega } \widetilde{C}_1 }{ 4C_{\rm eq} } $, and $\theta \equiv -\frac{\hbar}{d^2m\nu} \theta_0$. Accordingly with the same assumptions for derivation of Eq. (\ref{wtil3}), $\theta \approx 10^{-6} \theta_0$, and therefore is very small, but now we keep it since we will show that important features appear due to the effects of the coupling mediated by the NEMS vibration.
Hamiltonian (\ref{Hamilf}) shows the effective coupling between the two TLR modes 
mediated by the NEMS phonon number and allows 
two direct applications which will be discussed in what follows: 
(\textit{i})  the QND Measurement of NEMS Phonon Number; and (\textit{ii}) the generation of a tripartite entanglement involving the mechanical oscillator and the TLR fields, working as a probe of the quantum character of oscillation of the mechanical device. 

\section{QND Measurement of NEMS Phonon Number} 
As a first application we will develop a scheme for mean phonon number QND measurement \cite{walls2007quantum}
of the NEMS through a measurement carried out in one of the TLRs. 
%
Let us assume that an external drive $\mathcal{F}$ (resonant with  $\widetilde{\omega}$) is applied on the TLR-2. The effective interaction Hamiltonian (\ref{Hamilf}) together with this external drive is then given by
\begin{eqnarray*}
H_{I}&=&\hbar\left(\mathcal{F}^*a_{2}+\mathcal{F}a_{2}^{\dag}\right)+\hbar\left(\theta_{0}+\theta b^{\dag}b\right)\left(a_{1}^{\dag}a_{2}+a_{1}a_{2}^{\dag}\right).
\end{eqnarray*}
The  quantum stochastic differential equations (QSDEs) governing the evolution of $a_1$ and $a_2$ are
\begin{eqnarray}
\frac{da_{1}}{dt}&=&-i\theta_{0}a_{2}-i\theta b^{\dag}ba_{2}-\frac{\kappa_{1}}{2}a_{1}+\sqrt{\kappa_{1}}a_{1in},
\label{mov.eq.1}
\end{eqnarray}
\begin{eqnarray}
\frac{da_{2}}{dt}&=&-i\theta_{0}a_{1}-i\theta b^{\dag}ba_{1}-\frac{\kappa_{2}}{2}a_{2}-i\mathcal{F}+\sqrt{\kappa_{2}}a_{2in},
\label{mov.eq.2}
\end{eqnarray}
where for $j=1,2$, $\kappa_j$  is the relaxation rate, and $a_{jin}$ is the corresponding noise operator induced by individual reservoirs for the radiation mode at TLR-$j$.
In the limit where $\theta_0/\kappa_{2},\theta/\kappa_{2}<<1$ the radiation mode in the TLR-2 relaxes to a stationary coherent state due to the driving field. However it affects the radiation mode in the TLR-1 as given by Eq. (\ref{mov.eq.1}) contributing with an additional noise term. In that situation the first two terms in the second member of Eq. (\ref{mov.eq.2}) can be neglected. The steady state of the radiation mode in TLR-2 is then given by
\begin{eqnarray*}
\left\langle a_{2}\right\rangle \approx \frac{-2i\mathcal{F}}{\kappa_{2}}\equiv\alpha_{2}.
\end{eqnarray*}
We assume (without loss of generality) a purely imaginary driving field $\mathcal{F}$, so that $\alpha_2$ is real.
We now take into account the residual effect of ($\theta_{0}/\kappa_{2}$) as an additional dissipative channel for TLR-1. In that situation, the QSDE for $a_{1}$, Eq.  (\ref{mov.eq.1}), then becomes
\begin{equation}
\frac{da_{1}}{dt}=-i\theta\alpha_{2}b^{\dag}b-\frac{\kappa_{1}}{2}a_{1}-\frac{\Gamma}{2}a_{1}+\sqrt{\kappa_{1}}a_{1in}+\sqrt{\Gamma}\tilde{a}_{1in},
\label{diferentialeq.adiabat}
\end{equation}
where $\Gamma=2\theta_{0}^{2}/\kappa_{2}$, and $\tilde{a}_{1in}$ is an additive quantum noise term.
Eq.(\ref{diferentialeq.adiabat}) can be exactly solved to give
\begin{eqnarray}
a_{1}(t)&=&a_{1}(0)e^{-\frac{\Gamma+\kappa_{1}}{2}t}-\frac{2i\alpha_{2}\theta b^{\dag}b}{\Gamma+\kappa_{1}}\left(1-e^{-\frac{\Gamma+\kappa_{1}}{2}t}\right)\nonumber\\
&&+\sqrt{\kappa_{1}}\int_{0}^{t}e^{\frac{\kappa_{1}+\Gamma}{2}(t^{'}-t)}a_{1in}(t^{'})dt^{'}\nonumber\\
&&+\sqrt{\Gamma}\int_{0}^{t}e^{\frac{\kappa_{1}+\Gamma}{2}(t^{'}-t)}\tilde{a}_{1in}(t^{'})dt^{'}.
\label{solvedeq.addiabat}
\end{eqnarray}

What is mostly relevant in this last equation is the contribution of the TLR-2 radiation field amplitude $\alpha_2$ to the coupling to the mechanical mode. In fact even though $\theta$ is very small compared to $\theta_0$ the stationary coherent field $\alpha_2$ can be made strong enough (through the driving field) to amplify the interaction between the quantum radiation mode in TLR-1 and the mechanical mode. This coupling can indeed be explored to give a measurable experimental quantity. For example the average 
photocurrent in the TLR-1, defined as $\left\langle I_{1}(t)\right\rangle=
i \sqrt{{\hbar\tilde{\omega}}/{2L}} {\left\langle a_{1}^{\dag}-a_{1}\right\rangle}$, can be calculated to give (assuming $\left\langle a_{1}(0)\right\rangle=0$, without loss of generality)
\begin{eqnarray}
\left\langle I_{1}(t)\right\rangle& = & -\frac{\alpha_{2}\theta\sqrt{8\hbar\tilde{\omega}}}{\sqrt{L}(\Gamma+\kappa_{1})} n_b\left(1-e^{-\frac{\Gamma+\kappa_{1}}{2}t}\right)
\label{7}
\end{eqnarray}
with $n_{b}=\left\langle b^{\dag}b\right\rangle$.
\begin{figure}[ht]
\includegraphics[scale=0.28, trim = 0 0 0 0]{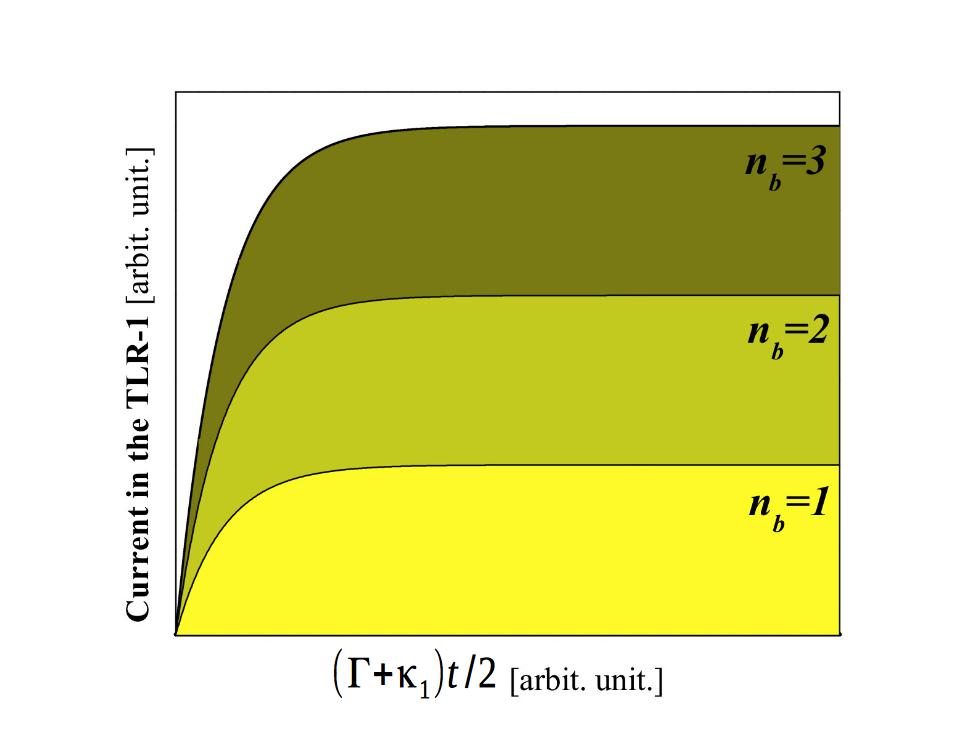}
\caption{ 
The photonic current $\left\langle I_{1}\right\rangle(t)$ in the TLR-1, 
as given by Eq.(\ref{7}) for three values of the NEMS Fock states with average phonon number $n_{b}=1,2,3$. 
For $t >> 2/(k_{1}+\Gamma)$, all three curves reaches a stationary threshold given by the phonon number $n_{b}$ in TLR-1 (remembering that $\theta<0$).}
\label{current}  
\end{figure}
To illustrate the TLR-2 photonic current profile in Fig. (\ref{current}) we plot $\left\langle I_{1}(t)\right\rangle$ by varying the NEMS average phonon number. It is clearly seen that the average number of phonons in the NEMS produces distinguishable values for the saturation of the photonic current. On the other hand it is immediate to see that the current variance, $\left\langle (\Delta I_1)^{2}\right\rangle= \left\langle I_{1}^2\right\rangle-\left\langle I_{1}\right\rangle^2$ is proportional to the variance of the phonon number, $\left\langle (\Delta b^{\dag}b)^{2}\right\rangle$. Therefore it can be used to infer the number statistics of the mechanical system. This subject is going to be considered further elsewhere.

\section{Entanglement Generation}

%
Now we analyze the distribution of entanglement in our tripartite system,  
as governed by the full quantized Hamiltonian (\ref{Hamilf}). Since $b^{\dag}b$ is a conserved quantity
we can solve the Heisenberg equations of motion for the operators $a_{1}$ and $a_{2}$ to obtain
%
\be
 a_k(t) = \sqrt{1-{\cal T}} a_{k}(0)-i\sqrt{{\mathcal{T}}} a_{j}(0), \,\,\, {j \ne k},
\ee
for  $j,k=1,2$.  
This is exactly the equation for a beam-splitter with intensity dependent transmittance \cite{walls2007quantum}
${\mathcal{T}} = \sin^2 [(\theta_0+\theta \langle b^{\dag}b\rangle t)]$. As is well known a beam splitter entangles only bosonic fields which are non-classical in the quantum optical sense \cite{Kim2002,Xiang-bin2002,DeOliveira2004}, i.e., when at least one of the individual inputs is described by a negative Glauber-Sudarshan P-distribution. For that reason the term in ${\mathcal{T}}$ depending on $\theta_0$ does not entangle the two radiation modes if they are initially in a classical state and therefore can be neglected for simplicity. The transmittance term dependent on the NEMS phonon number, however, allows the three modes, the mechanical and the photonic ones to be entangled, even when their initial state are classical. To illustrate this feature, let us consider that the system is prepared in a triple product of coherent states 
for the NEMS (index \textit{N}) and for the two TLR (indexes {1},{ 2}) as follows  
\be| \psi(0) \rangle = |\alpha \rangle_{N} |\beta \rangle_{ 1} |\gamma\rangle_{ 2}. \ee
The evolution will generate the conditioned entangled state 
\be
| \psi(t) \rangle = \sum_{n=0}^{\infty} C_n \,  
| n \rangle_{ N}  |\beta_n (t) \rangle_{ 1} |\gamma_n(t)\rangle_{ 2},    \label{CE3}
\ee
where $|n \rangle $ are the Fock state, with
$C_n = { e}^{-\frac12 |\alpha|^2} { \alpha^n }/{\sqrt{ n !}}$,
and 
\begin{eqnarray}  \label{CE4}
\beta_n (t)  & = &    \beta \, \cos(n \theta t) - i \gamma \, \sin( n\theta t) \nonumber \\
\gamma_n (t) & = &   \gamma \, \cos(n \theta t) - i \beta  \, \sin( n\theta t).         
\end{eqnarray}
For example, considering  $\theta t = \pi $, the state in Eq.(\ref{CE3}) becomes
\be
|\psi(t)\rangle=
|\alpha_+\rangle_{ N} |\beta\rangle_{ 1}|\gamma\rangle_{ 2} + 
|\alpha_-\rangle_{ N} |-\beta\rangle_{ 1}|-\gamma\rangle_{ 2},\label{CE5}
\ee
being $|\alpha_{\pm}\rangle \propto (|\alpha\rangle\pm|-\alpha\rangle)$ the even (+) and odd (-) coherent states. So, not only the three modes are entangled, but the NEMS is also left in a  mesoscopic cat state conditioned on the state on the TLR modes.
 
\begin{figure}[ht]
\includegraphics[scale=0.25]{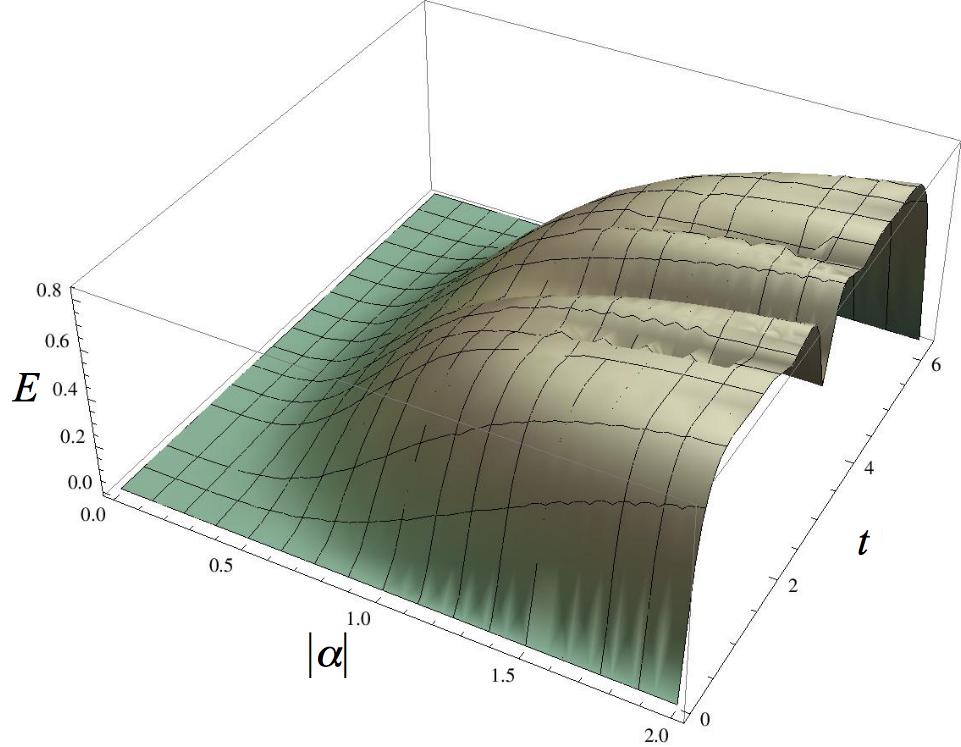}
\caption{
Linear entropy of the partition ${ N|12}$ quantifying the entanglement 
between the NEMS and the two TLRs as function of time and $| \alpha |$. 
We choose $\alpha = \beta = \gamma \in \mathbb{R}$. 
Note the entanglement's recurrence due to the $2\pi$-periodic 
behavior of the functions in (\ref{CE4}).}
\label{fig1}
\end{figure}

\begin{figure}[ht]
\includegraphics[scale=0.8]{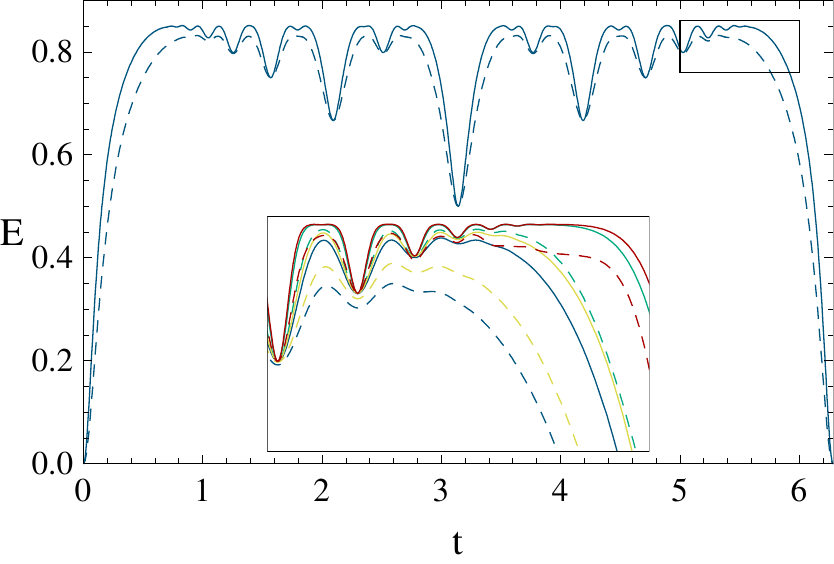}
\caption{Linear entropy of the partitions ${ N|12}$ (continuous) and 
${ 1|N2}$ (dashed) for $\alpha = \beta = \gamma = 2$ as a function of time. 
Since $\beta = \gamma$, $E_{ 1|N2} = E_{ 2|N1}$. 
Inset: The same linear entropies for distinct values of 
the initial state for $\alpha = 2$ and 
$\beta = \gamma = 2 $ (blue);
$\beta = 3 ,\gamma =4 $ (green);
$\beta = \gamma = 1+2i$ (yellow);
$\beta = 3+4i,\gamma=1+2i $ (red).}
\label{fig2}
\end{figure}

In a general way, when tracing over the partition ${ N}$ in (\ref{CE3}), 
one can see that the reduced state 
$\hat \rho_{ 12}$ is separable, {\it i.e.}, 
the NEMS is not able to generate entanglement between 
the TLRs. 
However an easy inspection of (\ref{CE3}) or even (\ref{CE5}) shows that 
when tracing over ${ 1}$ or ${ 2}$ the remaining system 
is entangled due to the non-orthogonality of coherent states, being non-etangled only for large $\beta$ and $\gamma$. Following the classification of 
\cite{PhysRevLett.83.3562} this is a {\it 1-part separable state}. 
For pure global states, the universal measure of entanglement is 
the von Neuman entropy, or the purity as given by the linear entropy $ E_{ A|B} = 1 - {\rm Tr} \hat \rho_{ B}^2$. 
Here $\hat \rho_{ B} $ represents the reduced density operator after 
tracing over the part ${ A}$ of the total system with
$\hat \rho = | \psi(t) \rangle \! \langle \psi(t)|$.
The behavior of the bipartitions from (\ref{CE3}) are encoded 
in the following equations for linear entropies
\begin{eqnarray}
E_{ N|12} & = & 1 - \! \sum_{n,m = 0}^{\infty} \! |C_n|^2 |C_m|^2 
                   {e}^{-| \beta_n - \beta_m |^2 -| \gamma_n - \gamma_m |^2  },\label{en12} \\
E_{ 1|N2} & = & 1 - \! \sum_{n,m = 0}^{\infty} \! |C_n|^2 |C_m|^2 
                   {e}^{-| \beta_n - \beta_m |^2  },\\
E_{ 2|N1} & = & 1 - \! \sum_{n,m = 0}^{\infty} \! |C_n|^2 |C_m|^2 
                   {e}^{-| \gamma_n - \gamma_m |^2  }.
\end{eqnarray}
By the Poissonian nature of $|C_n|^2$ and boundedness of the exponentials,  
all above sums are convergent.
In figs. \ref{fig1} and \ref{fig2} 
we show $E_{ N|12}$ for some initial coherent states when 
the summation is realized over 30 terms --- the error in this 
truncation is of the order of $10^{-17}$ for all the plotted 
curves. 
As one can see, although the system recurs to a non-entangled state after a period, it is usually highly entangled. At $\theta t=l \pi$, with $l=1, 3, 5,\ldots$ the state is as in Eq. (\ref{CE5}). This is a simple 
scheme for tripartite entanglement  generation involving  mechanical and photonic modes in a 
continuous variables regime, which  as we demonstrated can be easily implemented in a circuit.

\section{Summary}
In conclusion, 
in this paper we have investigated the possibility to directly couple capacitively two TLRs and a NEMS. The important feature here is that by treating the three parts of the system quantum mechanically 
we could obtain an interaction Hamiltonian that couples
the NEMS phonon number to the charges of the TLRs.  We have considered two simple applications of this setup. 
Firstly, the circuit may be considered for QND detection of the average number of phonons in the NEMS by photonic currents at one of the TLRs. We have shown that depending on NEMS average phonon number the current at one of the TLRs will furnish a characteristic and distinguishable behavior. This is quite important for actual detection schemes, since it is independent in principle on the inherent frequencies in the system. A second application is devised as a beam-splitter conditioned to 
the NEMS phonon number and this is employed into a scheme for generation of tripartite entanglement 
for a continuous variables system. This kind of device, an intensity-conditioned beam splitter is inexistent for optical fields and could in principle be used for quantum information processing, or detection purposes. 

\acknowledgements{This work is supported in part 
by CAPES and by FAPESP, and CNPq through the National Institute for Science and 
Technology of Quantum Information and through tthe Research Center in Optics and Photonics (CePOF). GJM acknowledges the support of the Australian Research Council grant 
CE110001013. MCO acknowledges insightful comments and suggestions made by Matthew LaHaye. OPSN is grateful to
L. D. Machado, S. S. Coutinho and K. M. S. Garcez for helpful discussions.}
\appendix
\section{Derivation of Hamiltonian (\ref{hamil1})}
Considering the source voltage for the circuit in Fig. \ref{fig5}, 
$V_{L_1}+V_{C_1}+V_{C_L}=V_{L_2}+V_{C_2}+V_{C_R}$
we arrive at the equations for the charge
\br
L_1\frac{d^2}{dt^2}Q_1(t)+\frac{1}{C_1}Q_1(t)&=&-\frac{Q_L(t)}{C_L(t)}\label{dq1}\\
L_2\frac{d^2 }{dt^2}Q_2(t)+\frac{1}{C_2}Q_2(t)&=&-\frac{Q_R(t)}{C_R(t)}\label{dq2}.
\er

Considering the total current in the system, $ I(t) = I_1 + I_2 $, 
we obtain
\be 
I(t)=\frac{d}{dt}Q_1(t)+\frac{d}{dt}Q_2(t)=\frac{d}{dt}Q_L(t)+\frac{d}{dt}Q_R(t),\ee
or
\be\label{qlr}Q_1(t)+Q_2(t)=Q_L(t)+Q_R(t)+ K,\ee
where $K$ is a constant which we will assume as null, without any loss of generality. 
Moreover the voltage over the whole capacitor is given by 
\be V_{C_T}=V_{C_L}-V_{C_R}=\frac{Q_L(t)}{C_L(t)}-\frac{Q_R(t)}{C_R(t)}.
\label{vt}
\ee

Combining Eqs. (\ref{qlr}) with (\ref{vt}), we obtain
{\br Q_L(t)&=&\frac{C_T}{C_R(t)}\left(Q_1(t)+Q_2(t)\right)+C_TV_{C_T}(t)\label{ql}\\
    Q_R(t)&=&\frac{C_T}{C_L(t)}\left(Q_1(t)+Q_2(t)\right)-C_TV_{C_T}(t)\label{qr},\er}
where $C_T^{-1}=C_L^{-1}(t)+C_R^{-1}(t)= \frac{2d}{\epsilon_0A}=2C_{\rm eq}^{-1}$.

Inserting Eqs.(\ref{ql}) and (\ref{qr}) in Eqs. (\ref{dq1}) and (\ref{dq2}) we obtain
\begin{widetext}
{\br
L_1\frac{d^2}{dt^2}Q_1(t)+\left(\frac{1}{C_1}+\frac{C_T}{C_L(t)C_R(t)}\right)
Q_1(t)+\frac{C_T}{C_L(t)C_R(t)}Q_2(t)&=&- \frac{C_T}{C_L(t)}V_{C_T}(t)\label{dq11}\\
L_2\frac{d^2}{dt^2}Q_2(t)+\left(\frac{1}{C_2}+\frac{C_T}{C_L(t)C_R(t)}\right)Q_2(t)+
\frac{C_T}{C_L(t)C_R(t)}Q_1(t)&=&+ \frac{C_T}{C_R(t)}V_{C_T}(t)\label{dq22}.
\er}
\end{widetext}
 
 Now, using the definitions for $C_L(t)$, $C_R(t)$ (see the main text) 
 and $C_T$ given bellow Eq.(\ref{qr}): 
 
 \be \frac{C_T}{C_L(t)C_R(t)}=\frac{\left(d^2-x^2(t)\right)}{2d\epsilon_0A},\ee
 and \be \frac{C_T}{C_L(t)}=\frac{\left(d-x(t)\right)}{2d},
 \,\,\frac{C_T}{C_R(t)}=\frac{\left(d+x(t)\right)}{2d}. \ee
 By defining 
 \be\frac{1}{\widetilde{C}_i}\equiv \frac1{C_i}+\frac{\left(d^2-x^2(t)\right)}{2d\epsilon_0 A},\ee 
 the canonical variables in the circuit $P_{i}(t)=L_{i}\dot{Q}_{i}=L_{i}I_{i}(t)$.
  eqs. (\ref{dq11}) and (\ref{dq22}) can be written as
 \begin{widetext}
\br
\frac{d}{dt}Q_1(t)&=&\frac1{L_1}P_1(t)\\
\frac{d}{dt}P_1(t)&=&-\frac{1}{\widetilde{C}_1}Q_1(t)-
\frac{\left(d^2-x^2(t)\right)}{2d\epsilon_0 A}Q_2(t)- 
\frac{\left(d-x(t)\right)}{2d}V_{C_T}(t)\label{dq112}\\
\frac{d}{dt}Q_2(t)&=&\frac1{L_2}P_2(t)\\
\frac{d}{dt}P_2(t)&=&-\frac{1}{\widetilde{C}_2}Q_2(t)-
\frac{\left(d^2-x^2(t)\right)}{2d\epsilon_0 A}Q_1(t)+ 
\frac{\left(d+x(t)\right)}{2d}V_{C_T}(t)\label{dq222}.
\er
Which, finally, allow us to derive the system Hamiltonian (\ref{hamil1}).
\end{widetext}

\section{Derivation of Hamiltonian (\ref{total})}

Two situations must be analyzed, when the NEMS behaves quantum-mechanically and when 
it behaves classically. Since the classical description can be derived from the quantum 
one we only describe the last one and latter recast the classical behavior. 
By writing $x(t)=\sqrt{\frac{\hbar}{2m\nu}}(be^{-i\nu t}+b^\dagger e^{i\nu t})$ we obtain
 $x^2(t)=\frac{\hbar}{2m\nu}(b^2e^{-i2\nu t}+(b^\dagger)^2 e^{i2\nu t}+2b^\dagger b+1)$. 
 For rapid oscillations ($\nu t>>1$),
$x(t)\approx 0$, and $x^2(t)\approx\frac{\hbar}{m\nu}(b^\dagger b+\frac12)\equiv x_{\rm rms}^2$. 
Thus
\be
\frac{1}{\widetilde{C}_i}= 
\frac1{C_i}+\frac{d^2-\frac{\hbar}{m\nu}(\langle b^\dagger b\rangle+\frac12)}{2d\epsilon_0 A}.
\label{wtil}
\ee 
Dividing both sides of Eq. (\ref{wtil}) by $L_i$  we obtain a relation between the two frequencies
  $\omega_i^2\equiv(L_iC_i)^{-1}$, and $\widetilde{\omega}_i^2\equiv(L_i\widetilde{C}_i)^{-1}$ 
  as 
  \be
  {\widetilde{\omega}_i^2}={\omega_i^2}+\frac{d^2- x_{\rm rms}^2}{2d\epsilon_0 AL_i}\label{wtil2},
  \ee 
  or
 \be
{\widetilde{\omega}_i^2}={\omega_i^2}+\frac{\omega_{i,\rm eq}^2}{2}\left(1-\frac{x_{\rm rms}^2}{d^2}\right),
 \ee 
 with $ \omega_{i,\rm eq}^2 \equiv ( C_{\rm eq} L_i)^{-1}$.  
We assume $d\approx 10^{-8}m$ $x^2_{\rm rms}/d^2=10^{-6}$ \cite{fabric,Ptoday} and so can be 
 disregarded from Eq. (\ref{wtil3}).
Typical experimental values settle $\omega_i=6$ GHz, and whenever  the second 
term in (\ref{wtil3}) is of the same order it should be taken into account. 
In any case in $\widetilde{\omega}_i^2$ we keep the maximal value the second term can take, 
i.e., we adopt $\widetilde{\omega}_i^2=\omega_i^2+\omega_{i,\rm eq}^2/2$, 
meaning that \be \frac{1}{\widetilde{C}_i}=\frac{1}{C_i}+\frac{1}{2C_{\rm eq}}.\ee

Thus the Hamiltonian (\ref{hamil1}) reduces to 
{
\br 
{\cal H} &\approx& \frac{P_1^2}{2L_1} + \frac{Q_1^2}{2\widetilde{C}_1} 
                                    + \frac{P_2^2}{2L_2} + \frac{Q_2^2}{2\widetilde{C}_2} 
\nonumber\\
&&+\frac{d^2-\frac{\hbar}{m\nu}(b^\dagger b+1/2)}{2d\epsilon_0 A}Q_1Q_2\nonumber\\
&&+\frac1{2}V_{C_T}(t)Q_1 -\frac12V_{C_T}(t)Q_2.\er}

Finally by assuming the conjugate variables (\ref{conjug}), 
which follow the standard commutation relation, 
we thus obtain the Hamiltonian (\ref{total}). 


\bibliographystyle{apsrev4-1}
\bibliography{nemsref}

\end{document}